\documentstyle[pra,aps,twocolumn]{revtex}
\begin{document}
\draft
\title{Entanglement, Information and Multiparticle Quantum Operations}
\author{Anthony Chefles}
\address{Department of Physical Sciences, University of Hertfordshire \\
       Hatfield AL10 9AB, Herts, UK \\ email: A.Chefles@herts.ac.uk}
\author{Claire R. Gilson}
\address{Department of Mathematics, University of Glasgow, Glasgow G12 8QQ, UK}
\author{Stephen M. Barnett}
\address{Department of Physics and Applied Physics, University of Strathclyde \\ Glasgow G4 0NG, UK}                                                                                                   \input epsf
\epsfverbosetrue \maketitle

\begin{abstract}
Collective operations on a network of spatially-separated quantum systems can be carried
out using local quantum (LQ) operations, classical communication (CC) and shared
entanglement (SE).  Such operations can also be used to communicate classical information
and establish entanglement between distant parties.  We show how these facts lead to
measures of the inseparability of quantum operations, and argue that a
maximally-inseparable operation on 2 qubits is the SWAP operation.  The generalisation of
our argument to $N$ qubit operations leads to the conclusion that permutation operations
are maximally-inseparable.  For even $N$, we find the minimum SE and CC resources which
are sufficient to perform an arbitrary collective operation.  These minimum resources are
$2(N-1)$ ebits and $4(N-1)$ bits, and these limits can be attained using a simple
teleportation-based protocol. We also obtain lower bounds on the minimum resources for the
odd case. For all $N{\geq}4$, we show that the SE/CC resources required to perform an
arbitrary operation are strictly greater than those that any operation can
establish/communicate.
 \\ *
\end{abstract}
\pacs{PACS numbers: 03.67.-a, 03.67.Hk}

\section{Introduction}
\renewcommand{\theequation}{1.\arabic{equation}}
\setcounter{equation}{0}

Many of the novel information-theoretic properties of quantum systems are attributable to
the existence of entanglement. Entanglement is responsible for the non-local correlations
which can exist between spatially separated quantum systems, as is revealed by the
violation of Bell's inequality\cite{Bell}.  It also lies at the heart of several
intriguing applications of quantum information, such as quantum
teleportation\cite{Teleport}, quantum computational speed-ups\cite{Shor,Grover} and
certain quantum cryptographic protocols\cite{Ekert}.

The central position of entanglement in quantum information theory, and its usefulness in
applications, has led to considerable efforts being devoted to finding a suitable measure
of how much entanglement a quantum system contains. This problem has been solved
completely for bipartite pure states\cite{Popescu}, and the accepted measure is the
subsystem von Neumann entropy, conventionally taken to the base 2, so that a
maximally-entangled state of a pair of two-level quantum systems, or {\em
qubits}\cite{Schumacher0}, possesses one unit of entanglement. This fundamental unit is
known as an {\em ebit}.

The production of entanglement requires the transmission of quantum information between
systems.  Conversely, the transmission of quantum information between systems can be used
to establish entanglement between them.  Perhaps the most perfect expression of this
duality is the fact that there are two equivalent definitions of the quantum capacity of a
communications channel\cite{Capacity1}. According to one definition\cite{Capacity2}, it is
the asymptotic maximum amount of quantum information that can be transmitted per use of
the channel, measured in qubits.  In the other\cite{Capacity3}, it is the asymptotic
maximum number of ebits of entanglement that can be established between the sending and
receiving stations, again per use of the channel. An important consequence of this
equivalence is the fact that no entanglement can be created without the transmission of
quantum information. That is, no entanglement can be created when only local quantum
operations are allowed, and only classical information can be transmitted.

Collective quantum operations involving multiple quantum systems can create entanglement
and be used to communicate classical information.  Conversely, the use of entanglement
shared by spatially-separated laboratories, in addition to facilities enabling classical
communication and arbitrary local quantum operations, permit these laboratories to carry
out collective operations upon a network of separated quantum systems. The ability to do
this will have interesting implications for many potential applications of quantum
information, such as distributed quantum computing, network quantum communication and the
production of novel multiparticle entangled states.

This paper extends the analysis presented in \cite{Small}, where we examined the
entanglement resources required to carry out collective quantum operations upon $N$
qubits, in particular, for the case of even {\em N}. In addition to giving a fuller
treatment of this problem, including an analysis of the odd case, we examine the classical
communication resources required to carry out an arbitrary collective operation upon $N$
qubits, and also the amount of classical information that such an operation can be used to
send.   An intriguing issue highlighted by these considerations is that of how we might
quantify the `inseparability' of a quantum operation, rather than that of a quantum state.
As we shall see, this inseparability has both classical and quantum aspects.

In section II, we examine the use of entanglement and classical communication to carry out
arbitrary collective operations upon a pair of qubits. A simple protocol for achieving
this, which uses quantum teleportation, is proposed.  Two classical and two quantum
measures of the inseparability of a quantum operation arise naturally from these
considerations. The quantum measures are analogous to the entanglement of
formation\cite{Formation} and distillation\cite{Distillation} of quantum states. These are
respectively the minimum amount of entanglement required to perform the operation, and the
maximum amount of entanglement that the operation can establish.  The classical measures
of inseparability are respectively the minimum amount of classical information required to
perform the operation, and the maximum amount of classical information that the operation
can be used to communicate. The relationship between these measures leads to the
conclusion that a maximally-inseparable quantum operation is the SWAP operation, or any
other which can be obtained from it by local unitary transformations.

The remainder of this paper is concerned with collective operations upon $N$ qubits.  The
particular issues we address are: how much bipartite entanglement can an operation be used
to establish and how much information can it be used to communicate?  Also, how much
bipartite entanglement and classical information are needed to perform an arbitrary
operation?

In section III, we develop a graph-theoretic framework for the representation of bipartite
entanglement and communication networks for $N$ laboratories.  Using this framework, we
generalise to the case of $N$ qubits our teleportation protocol.  We show that this
protocol is optimal in the class of protocols which operate by state teleportation.  We
also generalise our discussion of quantifying the inseparability of quantum operations to
the $N$-particle case.  As far as the `distillation' measures are concerned, which
quantifies the ability of a quantum operation to establish entanglement and communicate
classical information, we find that permutation operations are maximally-inseparable.
These operations can establish the largest amount of entanglement, and be used to
communicate the largest amount of classical information.

In section IV, we are concerned with minimising the entanglement and communication
resources required to perform an arbitrary quantum operation upon $N$ qubits.  There are
two distinct scenarios to consider here.  On the one hand, we may wish to determine the
minimum resources required to carry our an arbitrary operation just once.  We refer to
this as the `one-shot' scenario.  On the other hand, it may be the case that the $N$
laboratories share a very large amount of entanglement, and are able to communicate large
amounts of classical information.  They may wish to use these resources with maximum
efficiency to carry out an arbitrary operation many times.  The limit as both the
resources and the number of repetitions of the operation tends to infinity is known as the
{\em asymptotic} limit.  In this scenario, the asymptotically minimum resources are the
minimum entanglement and classical communication that must be used, on average, per run of
the operation.

We find that in terms of both entanglement and communication, our
teleportation protocol is optimal, in both the one-shot and
asymptotic scenarios, for even $N$.  We obtain lower bounds on the
minimum resources for the odd case.  We show that, for all
$N{\geq}4$, the classical communication and entanglement resources
required to carry out an arbitrary operation are strictly greater
than the amount of entanglement that can be established, and the
amount of classical information that can be sent, by any
particular operation.  We also show that if the manipulation of
these resources obeys the same efficiency restrictions as those
found in entanglement swapping\cite{Swapping} and indirect
communication, then the teleportation protocol is optimal for all
$N{\geq}12$, and for all $N{\geq}4$ for entanglement resources, in
the one shot case if only integer resources are allowed.

\section{Operations involving two qubits}
\renewcommand{\theequation}{2.\arabic{equation}}
\setcounter{equation}{0}

We consider first the simple case of just two qubits.  Suppose that two parties, by
convention Alice and Bob, occupy laboratories $A$ and $B$ which contain qubits ${\alpha}$
and ${\beta}$ respectively. The Hilbert spaces of these systems are denoted by ${\cal
H}_{\alpha}$ and ${\cal H}_{\beta}$, so that the Hilbert space of the collective system
${\alpha}{\beta}$ is the tensor product space ${\cal H}_{\alpha}{\otimes}{\cal
H}_{\beta}$.  In addition to these systems, Alice and Bob also possess auxiliary local
quantum systems, shared entanglement and a two-way classical communication channel. This
setup is illustrated in figure (1).   Using these resources, Alice and Bob can perform any
collective operation by carrying out the following four steps: \\ *

{\noindent}{\bf Step 1}: Alice teleports the state of ${\alpha}$ to Bob in laboratory $B$.
This costs 1 ebit of entanglement and 2 classical bits from $A$ to $B$. \\ *

{\noindent}{\bf Step 2}: Bob, possibly making use of his auxiliary systems, carries out
the operation locally upon the compound system. \\ *

{\noindent}{\bf Step 3}: Bob teleports the final state of Alice's qubit back to her.  This
costs 1 ebit of entanglement and 2 classical bits from $B$ to $A$. \\ *

{\noindent}{\bf Step 4}: (Selective operations only) Bob transmits
to Alice any classical information that he might have obtained at
the end of his LQ operation.  This step applies only to
(generalised) measurements, in which case it would be information
about the result. \\ *

Thus, the total CCSE resources required to perform an arbitrary collective operation on
${\alpha}{\beta}$ using teleportation, such that Alice and Bob share the same classical
information at the end, are
\begin{equation}
2\;{\mathrm ebits}+2\;{\mathrm bits}(A{\rightarrow}B)+2\;{\mathrm
bits}(B{\rightarrow}A)+C_{S}(B{\rightarrow}A).
\end{equation}

\begin{figure}

\epsfxsize7cm \centerline{\epsfbox{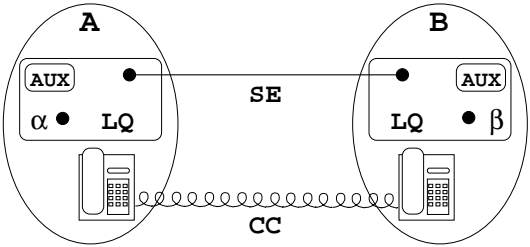}}

\vspace*{1cm}

\caption{Illustration of the experimental setup considered in section 2. Laboratories $A$
and $B$ contain respective qubits ${\alpha}$ and ${\beta}$.  Their aim is to perform an
arbitrary collective operation on these system using shared entanglement (SE) and a
two-way classical communication (CC) channel.  They are also able to perform arbitrary
local quantum (LQ) operations, possibly involving local auxiliary quantum systems and
their respective parts of the entangled systems. }

\end{figure}

The supplementary information $C_{S}(B{\rightarrow}A)$ is that
which is conveyed by Bob to Alice in step 4.  This additional
information will be created when the operation, represented by a
completely-positive, linear, trace-preserving map ${\cal L}$, is
selective. The most general kind of operation which gives rise to
non-zero supplementary information is a generalised measurement. A
generalised measurement with $M$ outcomes is described by $M$
positive, Hermitian operators $E_{r}$, where $r=1,{\ldots},M$ and
${\sum}_{r}E_{r}=1$. These operators form a positive,
operator-valued measure (POVM)\cite{Kraus} and each of them
corresponds to a distinct outcome.  If the initial state of
${\alpha}{\beta}$ is described by the density operator ${\rho}$,
then the probability $p_{r}$ of obtaining outcome $r$ is given by
${\mathrm Tr}{\rho}E_{r}$.  The supplementary information
generated at Bob's laboratory is given by the Shannon entropy of
this distribution
\begin{equation}
C_{S}=-\sum_{r=1}^{M}p_{r}{\log}_{2}p_{r}.
\end{equation}
This quantity can take on any non-negative real value. Clearly, it is zero when the
operation is non-selective. If, however, we consider an operation described by the POVM
\begin{equation}
E_{r}=\frac{1}{M},
\end{equation}
where Bob records the outcome, then the supplementary information is equal to
${\log}_{2}M$, which diverges as $M{\rightarrow}{\infty}$.  For this operation, one cannot
decrease the supplementary information using any information that Alice may have about the
initial state ${\rho}$, since the probability distribution is uniform regardless of what
the initial state is.

For selective operations, the transmission of this supplementary
information will have epistemological significance for Alice which
may be important in some applications.  She may, for example, wish
to carry out some local operation upon her subsystem, depending on
the supplementary information she receives from Bob.  For the
remainder of this paper however, we shall not be concerned with
$C_{S}$, and when we speak of the classical information required
to complete a quantum operation, we will mean that which is needed
to carry it out non-selectively.  In this paper, we shall be
concerned largely with unitary operations anyway, which are
non-selective.

Returning to the teleportation protocol, it may be the case that the CC and SE resources
required to perform a {\em particular} operation, ${\cal L}$, are less than those required
to perform {\em any} operation, by this method. Let us denote by $C_{R}({\cal
L}:A{\rightarrow}B)$, $C_{R}({\cal L}:B{\rightarrow}A)$ and $E_{R}({\cal L})$ the number
of classical bits transmitted in each direction and number of ebits of entanglement
required to carry out ${\cal L}$. These may be regarded respectively as classical and
quantum measures of how nonlocal the operation is, and $E_{R}({\cal L})$ is therefore
somewhat analogous to the entanglement of formation of quantum states\cite{Formation}.

Alternative classical and quantum measures of inseparability arise naturally if we
consider the fact that collective operations on quantum systems can be used to transmit
classical information and establish entanglement between distant locations.  Let us define
the quantities $C_{C}({\cal L}:A{\rightarrow}B)$, $C_{C}({\cal L}:B{\rightarrow}A)$ and
$E_{C}({\cal L})$, respectively the maximum number of classical bits that the operation
can be used to communicate in each direction, and the maximum number of ebits of
entanglement that it can create between $A$ and $B$.  $E_{C}({\cal L})$ is correspondingly
analogous to the entanglement of distillation of quantum states\cite{Distillation}.   We
must have
\begin{eqnarray}
C_{C}({\cal L}:A{\rightarrow}B)&{\le}&C_{R}({\cal L}:A{\rightarrow}B), \\ * C_{C}({\cal
L}:B{\rightarrow}A)&{\le}&C_{R}({\cal L}:B{\rightarrow}A), \\ * E_{C}({\cal
L})&{\le}&E_{R}({\cal L}).
\end{eqnarray}
The first two inequalities come from the fact that all classical information that the
operation can be used to transmit must, in figure (1), be sent over the classical channel.
Equivalently, no classical information can be transmitted using LQSE operations alone.
Were this not the case, it would be possible to violate relativistic casuality.  An
intriguing argument for this has recently been described by Eisert {\em et
al}\cite{Eisert}.  The third inequality comes from the fact that entanglement cannot
increase under LQCC operations. For one-way classical communication, this has been shown
by Horodecki and Horodecki\cite{Horodecki}, to be also equivalent to the impossibility of
superluminal communication.

As a consequence of the teleportation protocol, the minimum CCSE resources required to
perform any particular operation will not exceed  2 ebits of entanglement and 2 classical
bits each way. The most nonlocal quantum operations with regard to the resource measures
$E_{R}$ and $C_{R}$ are those for which the minimum values of these quantities are both
equal to 2.  Inequalities (2.4)-(2.6) imply that the maximum values of the $E_{C}$ and
$C_{C}$ cannot exceed 2.  Any operation which saturates the limits of 2 on the latter
measures must also then saturate inequalities (2.4-2.6), and can be termed a
maximally-inseparable operation.

One such operation is the SWAP operation.  This is a unitary operation $U_{S}$ which, for
any state $|{\psi}_{\alpha}{\rangle}{\in}{\cal H}_{\alpha}$ and any state
$|{\psi}_{\beta}{\rangle}{\in}{\cal H}_{\beta}$, acts as follows:
\begin{equation}
U_{S}|{\psi}_{\alpha}{\rangle}{\otimes}|{\psi}_{\beta}{\rangle}=|{\psi}_{\beta}{\rangle}{\otimes}|{\psi}_{\alpha}{\rangle},
\end{equation}
that is, it exchanges the states of ${\alpha}$ and ${\beta}$. The
ability of SWAP to create 2 ebits of entanglement and transmit 2
classical bits each way is easily demonstrated.  We shall now do
this, with reference to figures (2) and (3).  The remarkable
properties of the SWAP operation are also described by Collins
{\em et al}\cite{Collins} and Eisert {\em et al}\cite{Eisert}.

\begin{figure}[tbh]

\epsfxsize5cm \centerline{\epsfbox{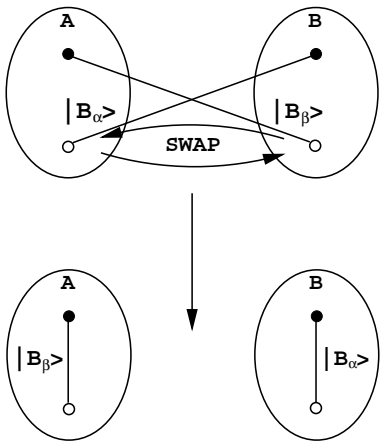}}

\vspace*{1cm}

\caption{Illustration of how the SWAP operation can be used to communicate two classical
bits each way between Alice and Bob.}

\end{figure}
In figure (2), Alice and Bob initially share 2 ebits of entanglement in the form of Bell
states\cite{Me}.  Using superdense quantum coding\cite{Coding}, Alice and Bob can each
manipulate one of their particles, those represented by hollow circles, to produce any of
the 4 Bell states that they wish. The final shared  Bell states are $|B_{\alpha}{\rangle}$
and $|B_{\beta}{\rangle}$.  The SWAP operation is them performed on the  states of the
hollow qubits, resulting in each party being in possession of the entire Bell state which
the other party created.  Each then performs a Bell measurement, which has 4 possible
outcomes and thus reveals 2 bits of information, showing how SWAP can transmit 2 classical
bits each way.

Figure (3) shows how SWAP can be used to establish 2 ebits of entanglement between Alice
and Bob.  Each party initially possesses 1 local ebit of entanglement.  If the SWAP
operation is used to interchange the states of one particle from each entangled pair, the
result is that Alice and Bob share 2 ebits of entanglement.

Notice that the SWAP operation cannot be used to create 2 ebits of entanglement, and
communicate 2 classical bits each way, simultaneously.  In fact, looking at figures (3)
and (4), we can see that one of these processes is essentially the time-reverse of the
other.

\begin{figure}

\epsfxsize5cm \centerline{\epsfbox{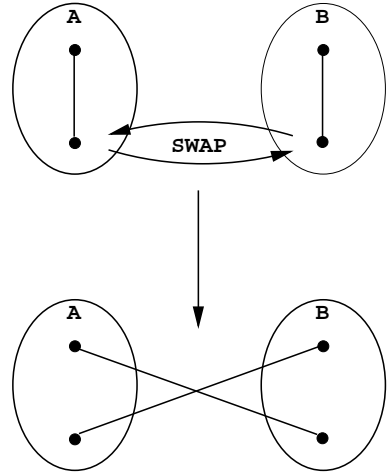}}

\vspace{1cm}

\caption{Illustration of how the SWAP operation can be used to establish 2 ebits of
entanglement between $A$ and $B$.}

\end{figure}

A broader class of maximally-inseparable operations on 2 qubits can be obtained by
considering those which are equivalent to SWAP up to a bilateral local unitary operation.
Specifically, any unitary operation $T$ of the form
$T=(U_{{\alpha}_{2}}{\otimes}U_{{\beta}_{2}})U_{S}(U_{{\alpha}_{1}}{\otimes}U_{{\beta}_{1}})$
must require the same entanglement and communication resources as $U_{S}$.  Here,
$U_{{\alpha}_{i}}$ and $U_{{\beta}_{i}}$ are local unitary operations on ${\alpha}$ and
${\beta}$ respectively.  The reason for this is simple: it is possible to convert this
operation into the SWAP operation by just local-unitary transformations, that is, without
any additional entanglement or classical communication resources.  This follows from the
simple observation that
$U_{S}=(U_{{\alpha}_{2}}^{\dagger}{\otimes}U_{{\beta}_{2}}^{\dagger})T(U_{{\alpha}_{1}}^{\dagger}{\otimes}U_{{\beta}_{1}}^{\dagger})$.

\section{Multiparticle systems, graphical representations and teleportation.}
\renewcommand{\theequation}{3.\arabic{equation}}
\setcounter{equation}{0}

Let us now extend our discussion to the case of $N$-particle systems.  Instead of just two
spatially-separated laboratories, we now have $N$ of them, which we label $A_{j}$, where
$j=1,{\ldots},N$.  In each of these laboratories is a qubit, and we label these $q_{j}$.
We are interested in the CCSE resources required to perform an arbitrary collective
quantum operation involving all $N$ qubits.

Each laboratory shares a certain number of ebits of entanglement
with every other laboratory.  In this paper, we shall, except
where indicated, take all entanglement to be in pure, bipartite
form.  The $N$ laboratories are also linked by classical
communication channels, so that each can communicate a certain
number of classical bits to the others. Each laboratory also
possesses auxiliary  quantum systems allowing arbitrary local
quantum operations to be performed.

The CCSE resources available to the network of laboratories are conveniently represented
using the concepts of graph theory\cite{Graph}. Recall that a graph $G=(V,E)$ is a set $V$
of vertices connected by edges comprising a set $E$.  If the edges have a sense of
direction indicating an asymmetrical relationship between the vertices it connects, the
graph is said to be a directed graph, or a digraph.  If there is no preferred direction,
the graph is undirected.

These resources can be represented by distinct entanglement and communication graphs. Both
graphs are comprised of $N$ vertices, each of which represents one of the laboratories
$A_{j}$.  The resource entanglement graph $G_{E}$ represents the amount of bipartite
entanglement shared between each pair of laboratories. Specifically, we write both the
$j$th laboratory and its corresponding vertex as $A_{j}$. The weight of the edge joining
vertices $A_{i}$ and $A_{j}$ is equal to the number of ebits of entanglement shared by
these laboratories.  The graph is characterised completely by the $N{\times}N$ resource
entanglement matrix ${\mathbf E}_{R}$.  The element $E_{R}^{ij}$ of this matrix is equal
to the number of ebits of entanglement shared by $A_{i}$ and $A_{j}$.  The diagonal
elements of this matrix are zero.

Clearly, ${\mathbf E}_{R}$ is symmetric and the graph $G_{E}$ is undirected.  These
observations follow from the fact that entanglement is a shared, rather than a directed
resource.

As an example, a resource entanglement graph for $N=4$ is depicted in figure (4).  This
corresponds to the following resource entanglement matrix:
\begin{equation}
{\mathbf E}_{R}=\left( \begin{array}{cccc} 0 & 3 & 2 & 6 \\ 3 & 0 & 1 & 0
\\ 2 & 1 & 0 & 0 \\ 6 & 0 & 0 & 0 \end{array} \right).
\end{equation}

Likewise, we can define a resource communication graph $G_{C}$. This represents the number
of classical bits that the laboratories can communicate directly to each other.  By
directly, we mean that it is not relayed by a set of intermediate laboratories from origin
to destination.  The weight of the edge running from $A_{i}$ and $A_{j}$ represents the
number of classical bits that $A_{i}$ can communicate directly to $A_{j}$. These weights
are the elements of a correspondingly defined resource communication matrix ${\mathbf
C}_{R}$.  The $ij$ element of this matrix, $C_{R}^{ij}$, is equal to the number of
classical bits that $A_{i}$ can communicate directly to $A_{j}$. The diagonal elements of
this matrix are also zero.  ${\mathbf C}_{R}$ is not necessarily symmetric and the graph
$G_{C}$ is directed, which follows from the fact that communication operations have a
natural sense of direction from sender to receiver.  An example of a resource
communication graph for $N=4$ is given in figure (5), which corresponds to the resource
communication matrix

\begin{equation}
{\mathbf C}_{R}=\left( \begin{array}{cccc} 0 & 1 & 4 & 0  \\ 2 & 0 & 0 & 9 \\ 0 & 0 & 0 &
0 \\ 5 & 0 & 0 & 0 \end{array} \right).
\end{equation}
\begin{figure}

\epsfxsize=5cm \centerline{\epsfbox{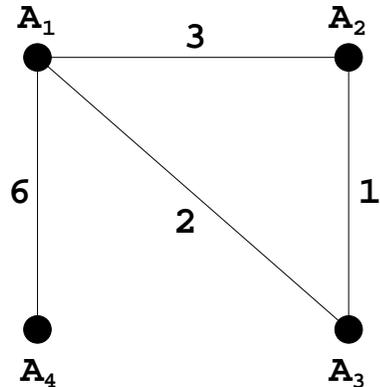}}

\caption{Example of an entanglement graph $G_{E}$ with $N=4$. This corresponds to the
resource entanglement matrix ${\mathbf E}_{R}$ in Eq. (3.1).}

\end{figure}

The fact that each pair of vertices may be joined by more than one edge means that $G_{C}$
is, strictly speaking, a multigraph, indeed a multidigraph since these edges are directed.
We do not, however, wish to unduly proliferate terminology, so we shall simply use the
term graph.

\begin{figure}

\epsfxsize5cm \centerline{\epsfbox{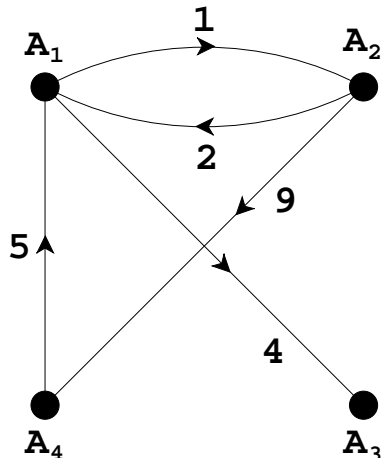}}

\caption{Example of a communication graph $G_{C}$ with $N=4$. This
corresponds to the resource communication matrix ${\mathbf C}_{R}$
in Eq. (3.2).}

\end{figure}

In either graph, an edge of weight zero is equivalent to no edge. Thus, if two vertices
are not linked by an edge in the graph $G_{E}$, then the corresponding laboratories share
no entanglement. Similarly, if there is no edge running from vertex $A_{i}$ to $A_{j}$ in
the graph $G_{C}$, then $A_{i}$ cannot communicate any classical information directly to
$A_{j}$.

Two quantities which will be of particular interest to us are the total shared
entanglement and the total number of classical bits that can be communicated.
Respectively, these are
\begin{eqnarray}
E_{R}&=&\frac{1}{2}\sum_{ij}E_{R}^{ij}, \\ * C_{R}&=&\sum_{ij}C_{R}^{ij}.
\end{eqnarray}
The factor of $1/2$ in Eq. (3.3) occurs as a consequence of the shared nature of
entanglement, which implies that the entanglement shared between each pair of laboratories
is counted twice in the summation.

Having established the framework within which we will work, let us now see how such
resources can be used to perform an arbitrary collective quantum operation upon the $N$
qubits $q_{j}$.  The teleportation-based procedure for 2 qubits described in the preceding
section admits a natural generalisation to the case of $N$ qubits, which we now describe.

We consider the situation in which all laboratories share entanglement and have the
resources for two-way classical communication with one particular laboratory.  Let this
laboratory be $A_{1}$.  It follows that the other laboratories can teleport the states of
their qubits to $A_{1}$.  The operation can then be carried out at $A_{1}$ as an LQ
operation.  The final states of the other qubits can then be teleported back to their
original laboratories, completing the procedure.

This multiparticle protocol generalises the first 3 steps of the 2-qubit protocol
described in the preceding section. It requires each of the laboratories
$A_{2},{\ldots},A_{N}$ to share 2 ebits of entanglement with $A_{1}$ and for 2 bits of
classical information to be communicated each way between each of them and $A_{1}$.  The
elements of the corresponding resource entanglement and communication matrices are
\begin{equation}
E_{R}^{ij}=C_{R}^{ij}=2|{\delta}_{i1}-{\delta}_{1j}|.
\end{equation}
The corresponding graphs $G_{E}$ and $G_{C}$ are depicted in figures (6) and (7).  The
total resource entanglement and communication are
\begin{equation}
E_{R}=\frac{C_{R}}{2}=2(N-1).
\end{equation}
The graph $G_{E}$ representing the entanglement resources required
by the teleportation protocol is said to be a {\em tree}.
Generally speaking, a tree is an connected, acyclic graph, that
is, one where every pair of vertices is connected by at least one
path, and where there are no closed paths.

Any quantum operation on $N$ qubits can be performed using this
method and thus, at least for the topology of entanglement and
communication in our protocol, the values of $E_{R}$ and $C_{R}$
in Eq. (3.6) are sufficient.

\begin{figure}

\epsfxsize6cm \centerline{\epsfbox{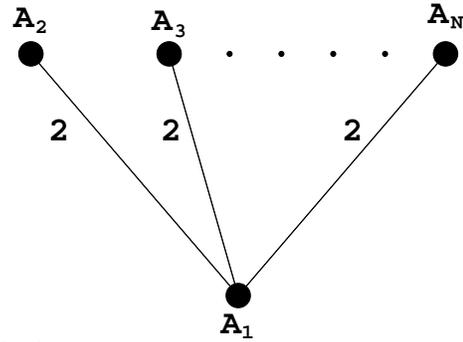}}

\caption{Resource entanglement graph for the teleportation protocol.}

\end{figure}

Much of the remainder of this paper will be concerned with the
issue of whether or not this protocol is optimal: that is, whether
or not there exists a procedure for carrying out any quantum
operation on $N$ qubits which less resources than this protocol.
Prior to doing so, it is of interest to determine whether or not
this protocol is the most efficient among those that operate by
teleportation of the states of the qubits concerned.

\begin{figure}

\epsfxsize6cm \centerline{\epsfbox{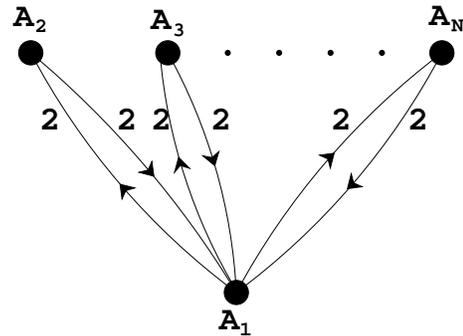}}

\caption{Resource communication graph for the teleportation protocol.}

\vspace*{1cm}

\end{figure}

We have $N$ laboratories $A_{i}$, each of which possesses a corresponding qubit $q_{i}$.
If we wish the $N$ laboratories to be able to carry out any collective operation upon the
$q_{i}$ by teleporting single qubits, then, as we now show, at least $2(N-1)$ such
teleportations must take place.

To see why, suppose that the first teleportation is from $A_{1}$ to $A_{2}$.  $A_{2}$ now
has information about $A_{1}$.  Secondly another lab $A_{r}$ teleports a state to $A_{3}$.
If they are completely different labs from the first pairs of laboratories then $A_{3}$
can hold information only about one other lab, $A_{r}$.  If, however, $r=2$, then $A_{3}$
can hold information about 3 qubits, $q_{1}$, $q_{2}$ and $q_{3}$.

The most efficient way to pass on information is for $A_{3}$ to teleport a state $A_{4}$
and so on. After $N-1$ steps the best possible situation is that one lab $A_{N}$  can have
information from all of the other labs. None of the other labs can have a complete set of
information. So now there must be at least a further $N-1$ communication events required
so that each of the first $N-1$ labs can get information from lab $A_{N}$. This gives a
total of a least $2(N-1)$ communications in all which costs $2(N-1)$ ebits.

We saw in the preceding section that the total resource entanglement for an arbitrary
operation upon two particles can be recovered if the operation in question is unitarily
equivalent to SWAP.  Also, for such an operation, the required classical communication
facilities required to complete an arbitrary operation can be fully used to communicate
useful information.  An important question is, does there exist an operation, or class of
operations which fulfills this role in the general $N$-particle case?

Let us denote the maximum total entanglement that can be established, and the maximum
number of classical bits that can be sent by any operation by $E_{C}$ and $C_{C}$
respectively.  To address this issue, it is helpful to partition the entire network of $N$
qubits into a single qubit and a compound system comprised of the remaining $N-1$ qubits.
How much entanglement can be established between the location of the isolated qubit and
the rest of the network?  Also, how much classical information can be transmitted in both
directions between the location of this qubit and the remainder?

In the teleportation protocol, a special status was given to laboratory $A_{1}$.  However,
this choice was arbitrary, and clearly this role could have been assumed by any
laboratory.  It follows that any collective quantum operation upon $N$ qubits can be
carried out with each laboratory sharing no more than 2 ebits of entanglement, and able to
exchange no more than 2 classical bits each way, with the rest of the network.  The
reasoning which led us to inequalities (2.4-2.6) then implies that no operation can be
used to establish more than 2 ebits of pure bipartite entanglement, or be used to exchange
more than 2 classical bits each way, between any particular laboratory and the rest of the
network.

The maximum total entanglement that can be established is then obtained by multiplication
of 2 ebits by the number of laboratories and then dividing by 2, since entanglement is
shared, giving
\begin{equation}
E_{C}{\le}N.
\end{equation}
The maximum number of classical bits that any collective operation
can be used to communicate is obtained by multiplying the maximum
amount of information that one laboratory can communicate, namely
2 bits, by {\em N}, the number of laboratories, giving
\begin{equation}
C_{C}{\le}2N.
\end{equation}
These bounds are tight, that is, they can be accessed by a
specific class of quantum operations, the permutation operations.

A unitary permutation operator upon $N$ qubits is described by
\begin{equation}
U_{P}|{\psi}_{1}{\rangle}{\otimes}{\ldots}{\otimes}|{\psi}_{N}{\rangle}=|{\psi}_{P(1)}{\rangle}{\otimes}{\ldots}{\otimes}|{\psi}_{P(N)}{\rangle},
\end{equation}
where $P(i)$ represents a permutation of the index $i{\in}[1,N]$. Here, we consider only
permutation operations which satisfy $P(i){\neq}i\;{\forall}i{\in}[1,N]$.

\begin{figure}[tbh]
\epsfxsize8cm \centerline{\epsfbox{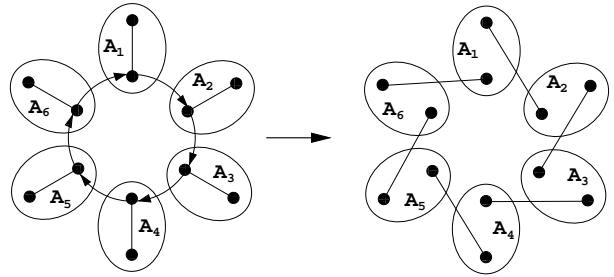}} \vspace*{1in}
\caption{Illustration of the production of $N$ ebits of
entanglement by the permutation operation.  Here, $N=6$ and the
permutation takes $\{1,2,3,4,5,6\}$ to $\{6,1,2,3,4,5\}$. One
qubit of each initial local ebit is transferred to the successive
laboratory, resulting in the final $N$ shared ebits.}
\end{figure}

To see that $N$ ebits of entanglement can be established using a permutation operation,
suppose that $A_{i}$ contains one local ebit, in the form of, for example, some standard
Bell state. We shall denote this state by $|B^{i,i}{\rangle}$. The first and second
indices denote the laboratories which possess the first and second qubits respectively.
Suppose now that the second qubits' states are permuted according to Eq. (3.9). This
transforms $|B^{i,i}{\rangle}$ into $|B^{i,P(i)}{\rangle}$. Following this permutation,
laboratories $A_{i}$ and $A_{P(i)}$ share the Bell state $|B^{i,P(i)}{\rangle}$.  There
are $N$ laboratories, and so $N$ shared ebits of entanglement in the form of Bell states
have been established.  This procedure is illustrated in figure (8).

To see that a permutation operation can be used to communicate
$2N$ classical bits, suppose that $A_{P^{-1}(i)}$ shares the Bell
state $|B^{i,P^{-1}(i)}{\rangle}$ with $A_{i}$. Locally, using
superdense coding, $A_{P^{-1}(i)}$ can manipulate the state of the
second qubit in this Bell state so that it becomes any of the four
possible Bell states. Figure (9) illustrates this scenario, where
each second qubit is represented by a hollow circle. We may
therefore write the state following this local manipulation as
$|B^{i,P^{-1}(i)}_{{\mu}(i)}{\rangle}$, where the integer
${\mu}(i){\in}[1,{\ldots},4]$. The permutation operation is then
carried out on the set of locally-manipulated qubits, resulting in
$A_{ i}$ being in possession of the state
$|B^{i,i}_{{\mu}(i)}{\rangle}$. By performing a Bell measurement,
$A_{i}$ can read the two bits of information sent by
$A_{P^{-1}(i)}$, and in total $2N$ bits have been communicated.

\begin{figure}[tbh]

\epsfxsize8cm \centerline{\epsfbox{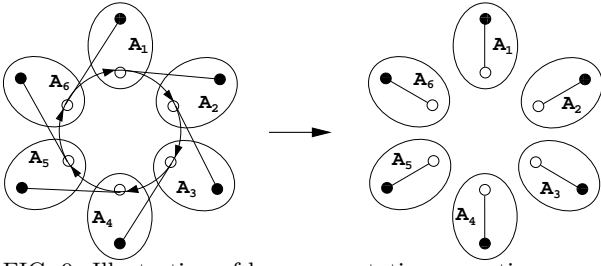}}

\caption{Illustration of how permutation operations can be used to
communicate $2N$ classical bits, using the same cyclic permutation
as in figure (8).  Each laboratory initially shares 1 ebit with
its successor, which it then manipulates into one of the 4 Bell
states.  The manipulated qubits are represented by hollow circles.
The permutation operation is then used to localise each of these
ebits in the successive laboratories. Individual local Bell
measurements are performed upon these, each of which reveals two
classical bits, or $2N$ bits in total.}

\vspace*{1cm}
\end{figure}

As is the case with the SWAP operation for 2 qubits, the number of ebits that $U_{P}$ can
establish is also the minimum amount of entanglement required to carry out this operation.
The same is true of the classical communication resources involved.  Suppose that $A_{i}$
shares one ebit of entanglement with $A_{P(i)}$ and can communicate 2 classical bits to
this location. Then the permutation operation can be carried out using these resources to
teleport the state of qubit $q_{i}$ from $A_{i}$ to $A_{P(i)}$.  Permutation operations,
including the SWAP operation, make maximally efficient use of the resources required to
carry them out.

As is also the case with the SWAP operation, any operation which is equivalent to $U_{P}$
up to an $N$-partite local unitary transformation, that is, any unitary operation $T$ of
the form
\begin{equation}
T=\left({\bigotimes}_{i=1}^{N}U^{2}_{i}\right)U_{P}\left({\bigotimes}_{j=1}^{N}U^{1}_{j}\right),
\end{equation}
where $U^{1}_{i}, U^{2}_{j}$ are arbitrary local unitary operations on qubits $q_{i}$ and
$q_{j}$, is also maximally-inseparable. This is a consequence of the fact that $U_{P}$ can
be obtained from $T$ by the local unitary operation
\begin{equation}
U_{P}=\left({\bigotimes}_{i=1}^{N}U^{2{\dagger}}_{i}\right)T\left({\bigotimes}_{j=1}^{N}U^{1{\dagger}}_{j}\right).
\end{equation}

Comparing (3.7) and (3.8) with (3.6), we see that the total amount of entanglement that
can be established, and the total amount of classical information that can be sent is
strictly less than that required to carry out an arbitrary operation using the
teleportation protocol, with the exception of the case $N=2$.  We have not, however,
established the optimality of the teleportation protocol. We examine this issue in the
following section.

\section{Resources Required to Perform Arbitrary Multiparticle Operations}
\renewcommand{\theequation}{4.\arabic{equation}}
\setcounter{equation}{0}

\subsection{Graph Symmetrisation}

The teleportation-based method for performing an arbitrary collective quantum operation
upon $N$ spatially separated qubits requires $E_{R}=2(N-1)$ ebits of entanglement and
$C_{R}=4(N-1)$ classical bits.  An obviously important question is: are these figures
optimal, in the sense that no less entanglement and communication will suffice?

Unlike the case of $N=2$, for general $N$ we cannot answer this
question by making use of the fact that the resource entanglement
and communication required by the teleportation protocol can
respectively be recovered and used to communicate messages, as can
be done with the SWAP operation.  For $N>2$, the values of $E_{R}$
and $C_{R}$ for the teleportation protocol, given by Eq. (3.6),
are strictly greater than the upper bounds on $E_{C}$ and $C_{C}$
in (3.7) and (3.8). Another approach must be taken to resolve this
issue. In this section, we show that, for even $N$, the resource
entanglement and communication required to perform an arbitrary
quantum operation upon $N$ qubits using the teleportation protocol
are indeed the minimum possible values. We describe a novel proof
technique, which we term {\em graph symmetrisation}, to establish
this fact.  The same method is then used to find lower bounds on
the minimum values of $E_{R}$ and $C_{R}$ for odd $N$. We find
that, for $N{\ge}4$, these lower bounds are strictly greater than
the upper bounds on $E_{C}$ and $C_{C}$ in Eqs. (3.7) and (3.8).

The problem we will investigate is the following.  A network of laboratories $A_{i}$
possesses shared bipartite entanglement, described by the graph $G_{E}$, and facilities
enabling limited classical communication between them, described by a graph $G_{C}$.  If
these graphs describe sufficient resources to enable any collective operation to be
performed upon their respective resident qubits $q_{j}$, then what lower bounds must the
corresponding values of $E_{R}$ and $C_{R}$ satisfy?

We commence by making the following observation: if the graphs $G_{E}(V)$ and $G_{C}(V)$
describe sufficient resources, then so does any other pair of graphs obtained from them by
a permutation of the vertices. Note that we have written the dependence of the graphs on
the vertex set explicitly here.  This makes sense intuitively. Nevertheless, here we
provide a short proof.  Let $G'_{E}(V)$ and $G'_{C}(V)$ be the entanglement and
communication graphs obtained from $G_{E}(V)$ and $G_{C}(V)$ by a permutation $P$ of the
vertex set. We may write $G'_{E}(V)=G_{E}(P[V])$ and $G'_{C}(V)=G_{C}(P[V])$, where $P[V]$
is the permutation. The reversibility of permutation operations implies that
$G_{E}(V)=G'_{E}(P^{-1}[V])$ and $G_{C}(V)=G'_{C}(P^{-1}[V])$. Consider now a quantum
operation ${\cal L}$ on the $N$ qubits. This also depends on the vertex set and so we
write it as ${\cal L}(V)$.  We can obtain another quantum operation ${\cal L}'$ from
${\cal L}$ by applying the same permutation to the vertex set, that is, ${\cal
L}'(V)={\cal L}(P[V])$, and ${\cal L}(V)={\cal L}'(P^{-1}[V])$.

If there exists an operation ${\cal L}'(V)$ which cannot be
performed using the resources described by the graphs $G'_{E}(V)$
and $G'_{C}(V)$, then by reversing the permutation $P$, it follows
that ${\cal L}(V)$ cannot be carried out using $G_{E}(V)$ and
$G_{C}(V)$, in contradiction with our premise. Thus, if the
resources described by the graphs $G_{E}(V)$ and $G_{C}(V)$ can be
used to carry out any quantum operation, then so do those
described by $G_{E}(P[V])$ and $G_{C}(P[V])$ for any permutation
{\em P} of the vertex set {\em V}.

Let us now consider the graphs ${\tilde G}_{E}$ and ${\tilde G}_{C}$, defined by
\begin{eqnarray}
{\tilde G}_{E}&=&\sum_{P[V]}G_{E}(V), \\ * {\tilde G}_{C}&=&\sum_{P[V]}G_{C}(V).
\end{eqnarray}
These graphs are constructed by summing over all of the graphs obtained from $G_{E}$ and
$G_{C}$ by permuting the vertices.  By summing, we mean summing the entanglement and
communication represented by the weights of the edges.  The resource entanglement and
communication matrices for these graphs are easily obtained.  Their elements are
\begin{eqnarray}
{\tilde E}^{ij}_{R}&=&\sum_{P[V]}E_{R}^{P(i),P(j)}, \\ * {\tilde
C}^{ij}_{R}&=&\sum_{P[V]}C_{R}^{P(i),P(j)}.
\end{eqnarray}
These graphs are regular and complete.  A complete graph is one where each pair of
vertices is joined by an edge.  In the case of the graph ${\tilde G}_{C}$, this means that
each pair of vertices is connected by an edge in each direction.  A regular graph is one
where all edges have the same weight.  In a network represented by these graphs, all pairs
of laboratories share the same amount of entanglement, and can communicate the same amount
of classical information, in both directions.

For the purposes of illustration, the graphs ${\tilde G}_{E}$ and ${\tilde G}_{C}$ are
shown in figures (10) and (11) corresponding to the particular graphs $G_{E}$ and $G_{C}$
in figures (4) and (5).

The regularity and completeness properties are easily proven, and follow immediately from
the fact that the graphs ${\tilde G}_{C}$ and ${\tilde G}_{E}$, being defined as sums over
all vertex permutations, are clearly themselves permutation invariant.

The total resource entanglement and communication for these graphs, ${\tilde E}_{R}$ and
${\tilde C}_{R}$, are easily evaluated in terms of the corresponding resources represented
by the original graphs $G_{E}$ and $G_{C}$.  Take the case of ${\tilde E}_{R}$: there are
$N!$ permutations of the vertex set, so ${\tilde G}_{E}$ describes $N!$ times as much
entanglement as $G_{E}$, that is
\begin{equation}
{\tilde E}_{R}=N!E_{R}.
\end{equation}
Similarly,
\begin{equation}
{\tilde C}_{R}=N!C_{R}.
\end{equation}

\begin{figure}

\epsfxsize=5cm

\epsfysize=5.33cm

\centerline{\epsffile{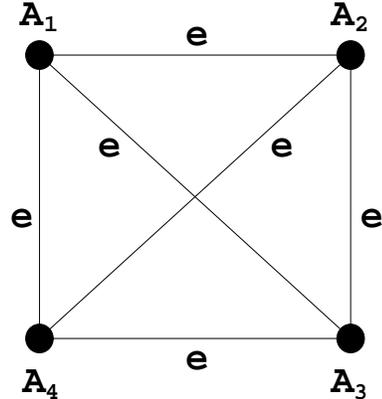}}

\caption{Symmetrised resource entanglement graph ${\tilde G}_{E}$ corresponding to the
graph $G_{E}$ in figure (4).  Here, $e$, which is given by eq. (4.7), is equal to 24.}

\end{figure}

\begin{figure}

\epsfxsize=6cm

\epsfysize=6.4cm

\centerline{\epsffile{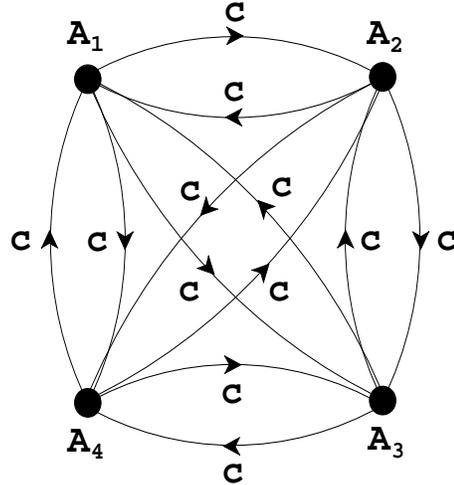}}

\caption{Symmetrised resource communication graph ${\tilde G}_{C}$ corresponding to the
graph $G_{C}$ in figure (5).  Here, $c$, which is given by Eq. (4.8), is equal to 42.}

\end{figure}

All edges in each of these graphs have the same weight, and it will be convenient to label
these two weights.  For ${\tilde G}_{E}$ and ${\tilde G}_{C}$, we denote these edge
weights simply by $e$ and $c$ respectively.  These are
\begin{eqnarray}
e&=&2(N-2)!E_{R}, \\ * c&=&(N-2)!C_{R},
\end{eqnarray}
which follows from Eqs. (4.5) and (4.6), and also from the fact
that the graphs ${\tilde G}_{E}$ and ${\tilde G}_{C}$ have
$N(N-1)/2$ and $N(N-1)$ edges respectively.  For the graphs in
figures (10) and (11), we find that $e=24$ and $c=42$.

There are $N!$ permutations of the vertex set.  The permutation invariance of the
sufficiency condition then implies that the resources represented by the graphs ${\tilde
G}_{E}$ and ${\tilde G}_{C}$ can be used to perform any operation $N!$ times.  By this, we
mean the following: suppose that $A_{i}$ contains $N!$ qubits. We can then define $N!$
sets of qubits, where each contains one qubit from each laboratory.  It will be possible
to perform the same operation separately upon every one of these sets.

In the next two subsections, we will use the formalism developed
here, together with inequalities (2.4-2.6) to establish lower
bounds on the values of {\em e} and {\em c}. These lead to lower
bounds on $E_{R}$ and $C_{R}$ through Eqs. (4.7) and (4.8).  We
shall treat the cases of even and odd {\em N} separately, since,
for even $N$, it is possible to use this technique to solve for
the minimum values of $E_{R}$ and $C_{R}$ which are sufficient to
carry out any operation.  These are those required to implement
the teleportation protocol described in section III.

\subsection{Necessary and Sufficient Resources for Even N}

Using the formalism we have set up, we can obtain the minimum values of $E_{R}$ and
$C_{R}$ exactly when $N$ is even.  The network of $N$ laboratories is assumed to possess
sufficient resources, described by the graphs ${\tilde G}_{E}$ and ${\tilde G}_{C}$, to
enable any operation to be carried out $N!$ times.  Here, we consider one particular
operation, which we will refer to as the pairwise-SWAP (PS) operation.  This operation has
the effect of swapping the state of a qubit at $A_{j}$ with that of one at $A_{j+1}$, for
all odd $j$.  If we write the two-particle SWAP operation exchanging the states of qubits
at $A_{j}$ and $A_{j+1}$ as $U_{S}^{j+1,j}$, then the PS operation may be written as
\begin{equation}
U_{PS}=U_{S}^{N,N-1}{\otimes}U_{S}^{N-2,N-3}{\otimes}{\ldots}{\otimes}U_{S}^{2,1}.
\end{equation}
This operation is illustrated in figure (12).

\begin{figure}[tbh]

\epsfxsize=5cm \centerline{\epsfbox{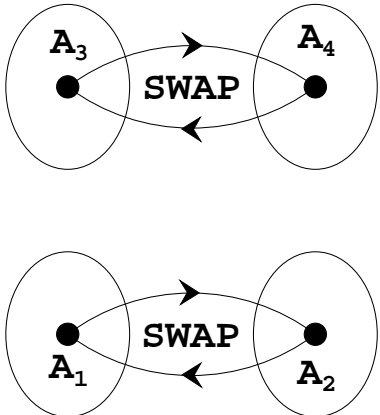}}

\caption{Depiction of the pairwise-SWAP (PS) operation for $N=4$.}

\vspace*{1cm}

\end{figure}

The PS operation is a permutation operation which leaves no vertex
invariant, and so it can be used to  establish $N$ ebits of
entanglement and to communicate $2N$ classical bits.  Performing
this operation $N!$ times can then be used to create $N!N$ ebits
of entanglement and to send $2N!N$ bits of classical information.
From the assumption that the graphs ${\tilde G}_{E}$ and ${\tilde
G}_{C}$ represent sufficient resources to carry out the $N!$-fold
PS operation, we can deduce the minimum values of $e$ and $c$, and
using Eqs. (4.7) and (4.8), those of $E_{R}$ and $C_{R}$, required
to do so.

To determine the minimum value of $e$, we will make use of the fact that entanglement
cannot increase under LQCC operations.  Consider the situation depicted in figure (13). We
partition the entire network into two sets. One contains the even laboratories
$A_{2},A_{4},{\ldots},A_{N}$, and the other contains the odd ones
$A_{1},A_{3},{\ldots},A_{N-1}$.  We shall refer to these sets as $S_{even}$ and $S_{odd}$.

\begin{figure}[tbp]
\epsfxsize6cm \centerline{\epsfbox{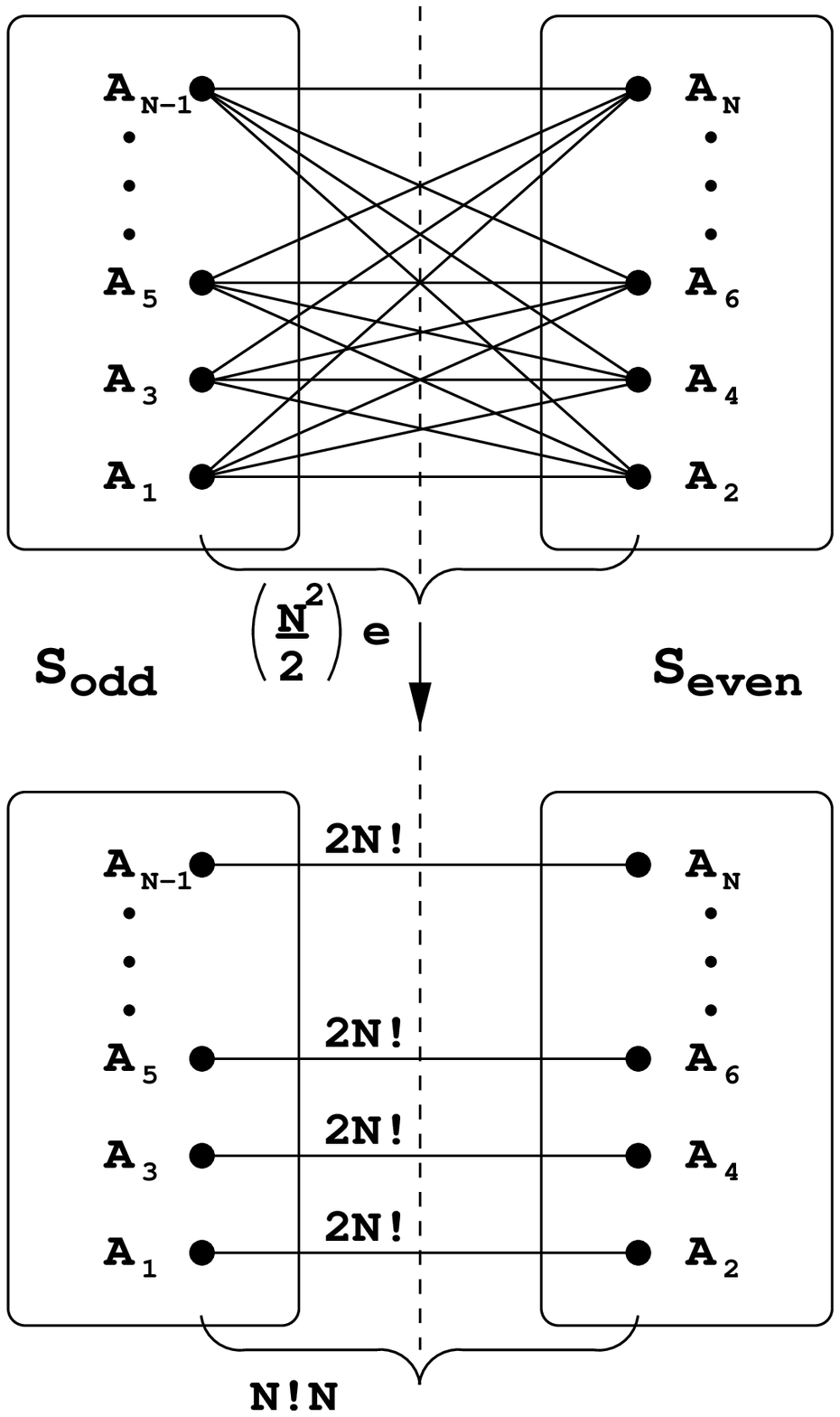}} \vspace*{1cm} \caption{Use of the resource
entanglement graph ${\tilde G}_{E}$ to carry out the $N!$-fold pairwise-SWAP operation.
Initially, the entanglement resources are distributed according to the graph ${\tilde
G}_{E}$. We have divided the $N$ laboratories into even and odd sets $S_{even}$ and
$S_{odd}$.  For the sake of clarity, we have not indicated the internal entanglement of
these sets. Each laboratory in $S_{odd}$ shares $e$ ebits of entanglement with each
laboratory in $S_{even}$.  These sets are separated by an imaginary partition, indicated
by the broken line.  Initially, these sets share $(N/2)^{2}e$ ebits, and the $N!$-fold PS
operation can create $N!N$ ebits.  The total entanglement shared across this partition
cannot increase, and the requirement that $e$ must be large enough to carry out the
$N!$-fold PS operation leads to inequalities (4.10) and (4.11). }

\end{figure}

The total entanglement initially shared by these sets can be calculated in a
straightforward manner. Each of the $N/2$ laboratories in $S_{odd}$ shares {\em e} ebits
with each laboratory in $S_{even}$, that is, $Ne/2$ ebits with $S_{even}$ in total. Adding
up the $N/2$ such contributions from the laboratories in $S_{odd}$ gives $(N/2)^{2}e$
ebits initially shared by $S_{even}$ and $S_{odd}$. The final entanglement they share is
$N!N$ ebits. The total entanglement that $S_{even}$ and $S_{odd}$ share cannot increase,
giving the inequality
\begin{equation}
\left(\frac{N}{2}\right)^{2}e{\geq}N!N.
\end{equation}
Making use of Eq. (4.7), we find that
\begin{equation}
E_{R}{\geq}2(N-1).
\end{equation}
This lower bound on the total resource entanglement is precisely the amount which is
required by the teleportation protocol.  Thus, for even $N$, the teleportation protocol is
optimal with regard to the required total resource entanglement.

This bound has been derived on the basis of the fact that, in a
multiparticle system, the (bipartite) entanglement shared by two
exhaustive subsets cannot increase under LQCC operations. Although
the entanglement initially shared by each pair of laboratories is
in pure, bipartite form, the transformation shown in figure (13)
may, at some point, manipulate the resource entanglement into,
possibly mixed, multiparticle entanglement. Our argument still
holds under these circumstances. If the final entanglement is in
multiparticle form, then in order to carry out the $N!$-fold PS
operation, $A_{j}$ and $A_{j+1}$ will have to be able to {\em
distill} $2N!$ ebits of pure, bipartite entanglement. The {\em
total} distillable entanglement between $S_{even}$ and $S_{odd}$
cannot increase, which leads to inequality (4.10) and thus the
teleportation bound in (4.11).

The nonincreasing of entanglement under LQCC operations is an asymptotic result.  It
follows that the teleportation protocol is asymptotically optimal for even $N$.  By
asymptotic\cite{Concentration}, we mean that, given a very large number of sets of
separated qubits, where the same, arbitrary operation is to be carried out on each set,
the teleportation protocol uses the minimum {\em average} entanglement that is required
per run of the operation.

In practical situations, it is often the resources required to carry out an operation
successfully just once that will be of interest.  For general information processing
tasks, the resources required in the `one-shot' scenario are at least equal to the
resources required asymptotically.  For the problem we have considered here, when $N$ is
even, the entanglement resources required in both scenarios are equal.  This is because
the teleportation protocol, which requires $2(N-1)$ ebits, can be used to carry out any
collective operation on $N$ qubits once.

The proof that the $N$ laboratories must also be able to send $4(N-1)$ classical bits
proceeds similarly.  The graph ${\tilde G}_{C}$ is assumed to represent sufficient CC
resources to perform any operation $N!$ times.  If this operation is the PS operation,
then it should then be able to communicate $2N!N$ bits. Given this, and the fact that each
laboratory can communicate $c$ classical bits to each other one, we can determine the
minimum value of $c$, from which we can infer the minimum of $C_{R}$ through Eq. (4.8).

Again, we partition the vertex set into $S_{even}$ and $S_{odd}$.
According to inequalities (2.4-2.5), the total amount of resource
communication between the sets $S_{even}$ and $S_{odd}$ cannot be
less than the amount of classical information that the $N!$-fold
PS operation can be used to communicate between these two sets.

According to ${\tilde G}_{C}$, each of the $N/2$ laboratories in
$S_{even}$ can communicate {\em c} classical bits to each one in
$S_{odd}$.  From this, we find that the maximum amount of
classical information that can be sent in either direction between
$S_{even}$ to $S_{odd}$ odd is $(N/2)^{2}c$ bits.

The $N!$-fold PS operation can be used to send $N!N$ bits in
either direction $S_{even}$ to $S_{odd}$. Inequalities (2.4-2.5)
imply that
\begin{equation}
\left(\frac{N}{2}\right)^{2}c{\geq}N!N.
\end{equation}
Making use of Eq. (4.8), we obtain

\begin{equation}
C_{R}{\geq}4(N-1),
\end{equation}
which is the amount of resource communication required to implement the teleportation
protocol.  We have thus shown that, in terms of the total resource entanglement and
communication, the teleportation protocol in section III is maximally efficient.

\subsection{Necessary Resources for Odd N}

Let us now examine the case of odd $N$.  We have been unable to find a specific operation
which proves that the minimum resources required to carry out any operation on an odd
number of qubits are those employed by the teleportation protocol.  However, using the
graph symmetrisation technique, it is still possible to obtain lower bounds on these
minimum resources.  As before, we assume that the graphs ${\tilde G}_{E}$ and ${\tilde
G}_{C}$ represent sufficient resources to perform any operation $N!$ times.  The specific
operation we shall consider here is the PS operation upon the first $N-3$ qubits, and a
separate, cyclic permutation of the remaining three.  For $N=3$, there is only this latter
part of the operation.   We shall refer to this as the PS+CP operation, and it is
illustrated in figure (14).

The PS+CP operation is again a permutation operation which leaves no vertex invariant.  It
follows that it can be used to establish $N$ ebits of entanglement and to communicate $2N$
classical bits.

We shall now apply the same arguments as those used for the PS operation for even $N$ to
obtain lower bounds on the resources required to carry out the PS+CP operation.  Again, we
divide the $N$ laboratories into two sets, $S_{even}$ and $S_{odd}$.

\begin{figure}[tbh]

\epsfxsize8cm \centerline{\epsfbox{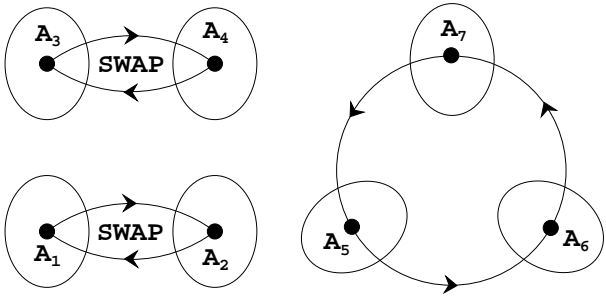}}

\caption{Depiction of the PS+CP operation for $N=7$}

\vspace*{1cm}

\end{figure}

Our aim, as before, is to obtain lower bounds on the minimum values of $c$ and $e$ from
the assumption that ${\tilde G}_{E}$ and ${\tilde G}_{C}$ represent sufficient resources
to perform this particular operation $N!$ times.

We begin by deriving a lower bound on the minimum sufficient resource entanglement
$E_{R}$.  In figure (15), the total initial entanglement between $S_{even}$ and $S_{odd}$
is depicted, as is the amount of entanglement that can be established by the $N!$-fold
PS+CP operation.  As before, these sets are divided by an imaginary partition, and the
total entanglement across this partition cannot increase.

Initially, each of the $(N-1)/2$ laboratories in $S_{even}$ shares {\em e} ebits with each
of the $(N+1)/2$ laboratories in $S_{odd}$.  The total amount of entanglement initially
shared by $S_{even}$ and $S_{odd}$ is then $(N^{2}-1)e/4$ ebits.  There are two
contributions to the amount of entanglement that can be created by the $N!$-fold PS+CP
operation. One is that created by the PS part of the operation on the qubits in the first
$N-3$ laboratories.  This can create $N!(N-3)$ ebits.  The second contribution comes from
the cyclic permutation on the remaining three laboratories.  This gives an additional
$2N!$ ebits.  Inequality (2.6) then implies
\begin{equation}
\left(\frac{N^{2}-1}{4}\right)e{\geq}N!(N-1).
\end{equation}
Making use of Eq. (4.8), we obtain a corresponding lower bound on $E_{R}$:
\begin{equation}
E_{R}{\geq}2\left(\frac{N}{N+1}\right)(N-1),
\end{equation}
that is, the teleportation bound multiplied by a factor of $N/(N+1)$.  The argument given
for even $N$ that this bound cannot be improved upon by converting the initial bipartite
resource entanglement into multiparticle entanglement also applies here.

\begin{figure}[tbp]
\epsfxsize6cm \centerline{\epsfbox{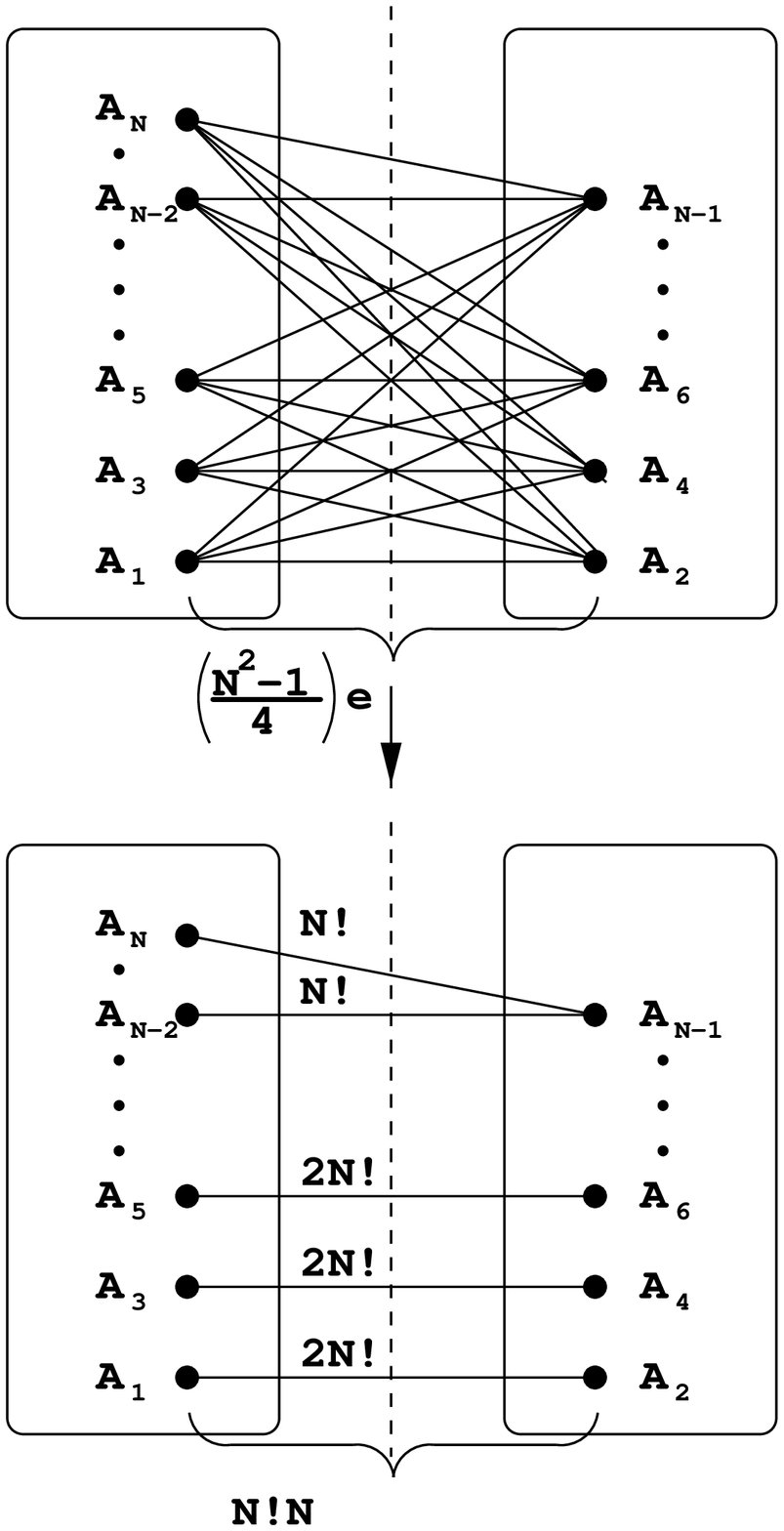}} \vspace*{1cm}
\caption{Use of the resource entanglement graph ${\tilde G}_{E}$
to carry out the $N!$-fold PS+CP operation. The initial
entanglement resources are distributed according to the graph
${\tilde G}_{E}$. Here, prior to carrying out the operation, the
sets $S_{even}$ and $S_{odd}$ share $(N^{2}-1)e/4$ ebits of
entanglement.  The $N!$-fold PS+CP operation can establish $N!N$
ebits between them. The total entanglement shared by these sets
cannot increase, and the requirement that $e$ must be large enough
to carry out the $N!$-fold PS+CP operation leads to inequalities
(4.14) and (4.15). }

\end{figure}

Let us now obtain a lower bound on the minimum resource
communication $C_{R}$.  As with the even case, we will make use of
the fact that the amount of classical information that the PS+CP
operation can be used to communicate, in either direction between
$S_{even}$ and $S_{odd}$, cannot exceed the amount of resource
communication in this direction that must be consumed in order to
implement the $N!$-fold PS+CP operation.

For the sake of concreteness, we shall consider communication from
$S_{even}$ to $S_{odd}$.  Initially, each of the $(N-1)/2$
laboratories in $S_{even}$ can communicate {\em c} classical bits
to each of the $(N+1)/2$ laboratories in $S_{odd}$.  This implies
that the total resource communication from $S_{even}$ to $S_{odd}$
is $(N^{2}-1)c/4$ bits.  It is easy to show that it is the same in
the opposite direction.

As with entanglement, the PS and CP parts of the $N!$-fold PS+CP
operation make distinct contributions to the amount of information
that this operation can be used to send from $S_{even}$ to
$S_{odd}$.  For a single implementation of PS+CP, the PS part can
communicate $(N-3)$ bits from $S_{even}$ to $S_{odd}$, while the
CP part can be used to send 2 bits.  Thus, the total amount of
classical information that the $N!$-fold PS+CP operation can be
used to send in either direction across the partition is $N!(N-1)$
bits.

The impossibility of this exceeding the resource communication in either direction across
the partition implies that
\begin{equation}
\left(\frac{N^{2}-1}{4}\right)c{\geq}N!(N-1).
\end{equation}
and, making use of Eq. (4.8), we obtain the corresponding bound for $C_{R}$:
\begin{equation}
C_{R}{\geq}4\left(\frac{N}{N+1}\right)(N-1),
\end{equation}
which, like the entanglement bound in (4.15), is the teleportation bound multiplied by
$N/(N+1)$.

Like the bounds in (4.11) and (4.13) for the even case, the lower
bounds we have obtained here for the minimum resource entanglement
and communication for odd {\em N} are asymptotic results. However,
the fact that the bounds (4.15) and (4.17) are not integers
suggests that if the available resources are at these bounds, they
may not be very useful in the one-shot case, where it is more
desirable to be able to transmit whole bits of classical
information, and to manipulate whole ebits of entanglement.  With
this in mind, let us return to the bound on $E_{R}$ in (4.15) and
consider the inequality
\begin{equation}
\frac{2N(N-1)}{N+1}=2(N-1)-2+\frac{4}{N+1}{\geq}2(N-1)-2
\end{equation}
where the equality is attained only in  the limit as $N{\rightarrow}{\infty}$.  If we are
to round this bound up to the next integer, we obtain $2(N-1)-1$.  Thus, the minimum
number of integer ebits able to carry out an arbitrary operation on an odd number of
qubits, in the one-shot case, is bounded from below by one ebit less than the
teleportation bound.

By a similar calculation, one can show, using the bound in (4.17),
that in the one-shot case, if classical information is to be
transmitted in integer amounts, then the minimum number of bits
needed to carry out an arbitrary operation on an odd number of
qubits is bounded from below by 3 bits less than the teleportation
bound. For $N=3,5$, a stronger bound of 2 bits less than resource
communication for the teleportation protocol is obtained.

With these observations in mind, the case of $N=3$ appears to be particularly significant.
For this case, in the one-shot scenario, we see that at least 3 ebits and 6 classical bits
are required.  However, we know that a permutation of 3 qubits can create 3 ebits or be
used to send 6 classical bits.  This implies that these bounds must also hold
asymptotically.

It is important to compare the bounds in (4.15) and (4.17), which
hold rigorously in both the one-shot and asymptotic scenarios,
with the maximum amount of entanglement that can be created, and
the maximum amount of classical information that can be sent, by
an $N$ qubit operation. With this in mind, we note the following
inequality, which holds for all $N{\geq}3$:
\begin{equation}
2\left(\frac{N}{N+1}\right)(N-1){\geq}N.
\end{equation}
The equality is obtained only when $N=3$.  This implies that, for all $N{\geq}4$, the
resources required to carry out an arbitrary operation exceed those that can be recovered,
either by re-establishing consumed entanglement, or using the resource communication which
was consumed to implement the operation to send useful messages.

As we saw in section II, this is not the case for $N=2$, which can be seen from the
properties of the SWAP operation.  The remaining case, that of $N=3$, is presently
unsolved.
\subsection{Transfer of Expendable Resources}

In our derivation of the lower bounds on the minimum resource entanglement and
communication needed to carry out any multiqubit operation, we used specific operations
where certain pairs of laboratories needed to be able to establish large amounts of
entanglement or communicate large amounts of classical information: more than is
represented by the corresponding edges in the graphs ${\tilde G}_{E}$ and ${\tilde
G}_{C}$.  Thus, to carry out either the $N!$-fold PS or PS+CP operation, the resources
from the other edges in these graphs must somehow be `transferred' to the edges which must
gain resources.

We can formalise this notion in the following way: consider a multiqubit operation ${\cal
L}$ on $N$ qubits.  If ${\cal L}$ is carried out $N!$ times, then depending on the initial
conditions, there may be some pairs of laboratories which will end up sharing more that
{\em e} ebits of entanglement, or exchanging more than {\em c} classical bits in either,
or perhaps both directions. Let is define the {\em target} entanglement and communication
graphs $G^{T}_{E}$ and $G^{T}_{C}$.  These will represent either the number of ebits
shared by each pair of laboratories or the number of classical bits communicated,
following the $N!$-fold implementation of ${\cal L}$. These graphs can be characterised by
the target entanglement and communication matrices ${\mathbf E}_{T}= \{E^{ij}_{T}\}$ and
${\mathbf C}_{T}= \{C^{ij}_{T}\}$, in the same way as for the resource graphs.

For ${\tilde G}_{E}$ and $G^{T}_{E}$, we will define a pair of complementary subsets of
the edge set, $S_{E+}$ and $S_{E-}$, in the following way: $S_{E+}$ is the subset of the
edge set, where each edge is denoted by the unordered pair $(i,j)$, such that
$E^{ij}_{T}>e$.  The set $S_{E-}$ contains all edges for which $E^{ij}_{T}{\leq}e$. These
sets contain the edges which respectively gain, and do not gain entanglement.

Similarly, for the classical communication graphs ${\tilde G}_{C}$ and $G^{T}_{C}$, we
will define the subsets $S_{C+}$ and $S_{C-}$ of the edge set.  $S_{C+}$ contains the
edges, represented by {\em ordered} pairs $[i,j]$, for which $C^{ij}_{T}>c$, and $S_{C-}$
contains all edges $[i,j]$ for which $C^{ij}_{T}{\leq}c$.

Here, we shall be particularly interested in the edges which gain resources.  In fact, the
resources contained in the other edges, contained in the sets $S_{E-}$ and $S_{C-}$, will
be considered {\em expendable}.  The total expendable entanglement and communication are
given by
\begin{eqnarray}
E_{E}&=&\frac{1}{2}\sum_{(i,j){\in}S_{E-}}{\tilde E}_{R}^{ij}, \\ *
C_{E}&=&\sum_{[i,j]{\in}S_{C-}}{\tilde C}_{R}^{ij}.
\end{eqnarray}

The question we would like to answer is: how much of the expendable entanglement or
communication can be transferred to the set $S_{E+}$ or $S_{C+}$?  We have been unable to
obtain the general solution to this problem, although the analysis of the PS operation
suggests intuitively appealing upper bounds.

For the $N!$-fold PS operation, the values of $E_{E}$ and $C_{E}$ are easily calculated,
where the sets $S_{E+}$ and $S_{C+}$ contain the edges linking laboratories whose qubits
are to be swapped.  We find that
\begin{eqnarray}
E_{E}&=&\frac{1}{2}(N^{2}-2N)e, \\ * C_{E}&=&(N^{2}-2N)c.
\end{eqnarray}
If each pair of swapped qubits generates 2 ebits of entanglement, then as we know, the
$N!$-fold PS operation can by used to create $N!N$ ebits.  The total amount of
entanglement which has been {\em added} to the set $S_{E+}$ is then $N!N-(Ne/2)$ ebits.
From inequality (4.10), we see that
\begin{equation}
N!N-\frac{Ne}{2}{\leq}\frac{E_{E}}{2}
\end{equation}
that is, at most half of the expendable entanglement can be added
to the edges in $S_{E+}$.  Whether or not this bound holds in
general for all $N$, and when the initial resource entanglement is
not described by a regular, complete graph, is currently unknown.
However, we can prove that it holds in general for $N=3$. Consider
3 laboratories, $A_{1}{\ldots}A_{3}$.  Let their initial and final
entanglement be described by the resource and target entanglement
graphs $G_{E}$ and $G_{E}^{T}$, characterised by the corresponding
matrices ${\mathbf E}_{R}$ and ${\mathbf E}_{T}$. The difference
between the initial and final entanglement between each pair of
laboratories can be represented by the matrix ${\mbox{\boldmath
${\Delta}$}}={\mathbf E}_{T}-{\mathbf E}_{R}$.

The fact that the total amount of entanglement shared by one laboratory and the other pair
cannot increase implies that the sum of the elements in each row or column of
${\mbox{\boldmath ${\Delta}$}}$ cannot exceed zero.  This also implies that the
entanglement between at most one pair of laboratories can increase.  Let this pair of
laboratories be $A_{1}$ and $A_{2}$.  Summing up the elements of ${\mbox{\boldmath
${\Delta}$}}$ in rows 1 and 2, together with the nonincreasing property of the column
sums, gives ${\Delta}_{12}{\leq}|{\Delta}_{13}+{\Delta}_{23}|/2$.  The numerator on the
right hand side is the total amount of entanglement lost.  We see that the entanglement
transferred to the edge (1,2) cannot exceed half of this loss.

This kind of entanglement loss was originally discovered in
association with {\em entanglement swapping}\cite{Swapping}.  It
would be useful to know whether or not it is an unavoidable
feature of all operations which transfer entanglement, and for an
arbitrary number of spatially separated systems. Any proof, or
disproof, of this conjecture must take into account the
possibility that the initial bipartite resource entanglement is
converted into multiparticle entangled states. Some progress has
recently been made towards developing a theory of conversion
between bipartite and multiparticle entangled states\cite{Linden}.
The study of certain particular situations has indicated that
these conversions are typically lossy. Consequently, we do not
believe that multiparticle entangled states will enable more
efficient entanglement transfer.

Returning to the $N!$-fold PS operation, we will show how a
similar result relating the expendable communication to the
communication that can be added to the edges in $S_{C+}$ can be
obtained.  The $N!$-fold PS operation can by used to send $2N!N$
bits. The total amount of communication which has been {\em added}
to the set $S_{C+}$ is then $2N!N-Nc$ bits. From inequality
(4.12), we see that
\begin{equation}
N!N-\frac{Nc}{2}{\leq}\frac{C_{E}}{2}
\end{equation}
that is, at most half of the expendable communication can be added
to the edges in $S_{C+}$.

This restriction holds in general if the expendable communication
is used to transmit information {\em indirectly} between pairs of
laboratories.  By `indirectly', we mean the following.  the weight
of an edge in  a resource communication graph is equal to the
number of bits that one party can transmit along a channel to some
other party, without passing through some intermediate laboratory.
Clearly, the sender can transmit more information to the receiver
if he sends some information via some intermediate laboratories.
By indirect communication, we mean this relaying procedure.

Thus, if the sender wishes to send ${\kappa}$ bits indirectly, he will use up at least
${\kappa}$ bits of resource communication sending this information to the intermediate
parties, who will in turn use up at least a further ${\kappa}$ bits of resource
communication relaying it to the receiver.  So, the ${\kappa}$ bits actually communicated
from sender to receiver cannot exceed the lower bound of 2${\kappa}$ bits depleted from
the resource communication.

For the remainder of this section, we will make the hypothesis
that at most half of the expendable resources can be transferred
is of general validity, and explore its consequences for odd {\em
N} and the PS+CP operation.   We will show that this leads to
tighter bounds than (4.15) and (4.17) on the amount of resource
entanglement and communication needed to carry out an arbitrary
operation on an odd number of qubits.

Let us return to the $N!$-fold PS+CP operation.  Beginning with entanglement, the amount
of entanglement transferred to the edges in $S_{E+}$ has two contributions.  The
entanglement transferred by the PS part of the operation is easily calculated to be
$(N-3)(2N!-e)/2$ ebits.  The amount transferred by the CP part is $3(N!-e)$ ebits.  The
expendable entanglement is that initially represented by all other edges in the graph
${\tilde G}_{E}$, and is found to be
\begin{equation}
E_{E}=\left[\frac{N^{2}}{2}-N-\frac{3}{2}\right]e.
\end{equation}
The assumption that at most half of the expendable entanglement can be transferred to the
edges in $S_{E+}$ leads to the inequality

\begin{equation}
N!N{\le}e\left[\frac{N-3}{2}+3\right]+\frac{E_{E}}{2}.
\end{equation}
From the relationship between {\em e} and the initial resource entanglement $E_{R}$,
expressed in Eq. (4.7), we find
\begin{equation}
E_{R}{\ge}\frac{2(N-1)}{1+3/{N^{2}}}.
\end{equation}

\begin{figure}

\epsfxsize9cm \centerline{\epsfbox{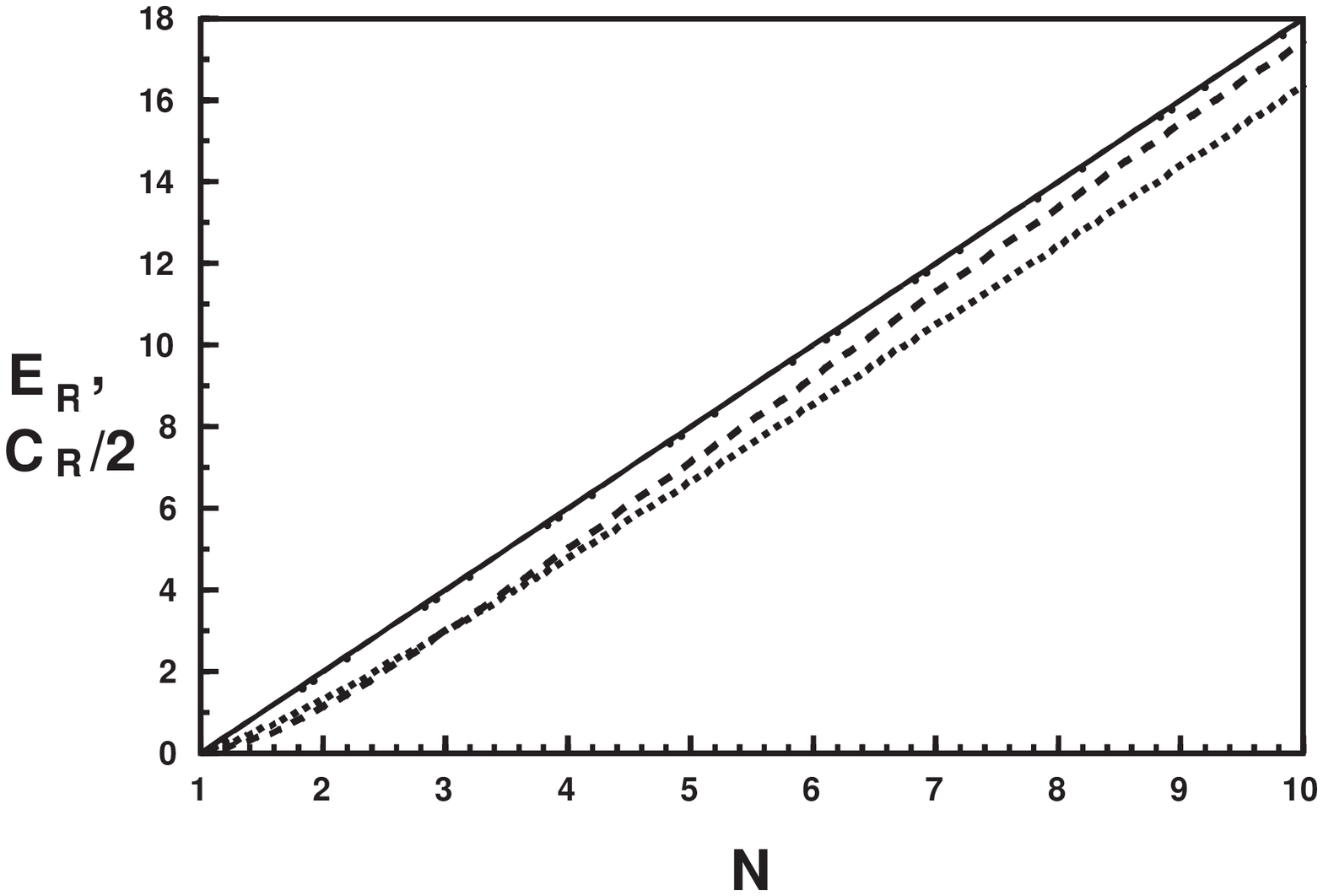}}

\caption{Lower bounds on the resource entanglement and communication versus the number of
qubits.  The solid line corresponds to the teleportation bound in (4.11) and (4.13). The
dotted line indicates the bounds in (4.15) and (4.17) for the PS+CP operation which hold
rigorously for odd N.  The dashed line corresponds to the bounds in (4.28) and (4.29), for
the PS+CP operation and odd N if at most half of the expendable resources can be
transferred. }

\end{figure}

Similar reasoning can be applied to the required minimum classical communication.  If at
most half of the expendable communication is transferable, then the minimum number of
classical bits required to perform the PS+CP operation with odd $N$ is bounded by

\begin{equation}
C_{R}{\ge}\frac{4(N-1)}{1+3/{N^{2}}}.
\end{equation}

Assuming that inequalities (4.28) and (4.29) hold, let us deduce the minimum integer
resources for the one shot case, as we did in the previous subsection.  To this end, we
note the inequality
\begin{equation}
\frac{2(N-1)}{1+3/{N^{2}}}{\ge}2(N-1)-1,
\end{equation}
for $N{\geq}3$, with the equality only being attained when $N=3$.
From this inequality, we see that for all $N{\geq}4$, the minimum
integer resource entanglement is equal to that required to
implement the teleportation protocol.  By a similar calculation,
one can show that for all $N>3$, the minimum integer resource
communication is at least equal to one bit short of the
teleportation bound, and that for all $N{\geq}12$, it is the
teleportation bound.

Figure (16) illustrates the main asymptotic bounds we have considered in this paper: the
teleportation bound, the bound derived from the PS+CP operation derived in the previous
subsection and the bound derived from the PS+CP operation based on the assumption that
only half of the expendable resources are transferable.

To summarise this subsection, we have worked on the assumption
that at most half of the expendable resources can be transferred.
The analysis of the PS operation supports this conjecture that it
is true for all $N$.  If it indeed is true in general, then in the
one-shot case, the teleportation protocol is optimal with regard
to the resource entanglement for all odd $N{\neq}3$, and also in
terms of the resource communication for all odd $N{\geq}13$, if
only integer resources are permitted.

\section{Discussion.}
\renewcommand{\theequation}{5.\arabic{equation}}
\setcounter{equation}{0}

In this paper, we have examined the properties of collective quantum operations performed
upon spatially separated quantum systems.  We have considered a network of $N$ spatially
separated laboratories, each of which contains one qubit.  The network is equipped with
facilities for classical communication and local quantum operations, and each pair of
laboratories also shares bipartite pure entanglement.

This scenario we have considered helps to emphasise the fact that the final state of each
system will depend upon the initial states of the others. The evolution thus requires
information to be exchanged between the systems.  In classical physics, this is simply
classical information.  If the systems are quantum mechanical, then the exchange of
quantum information is necessary.

The transmission of quantum information from one location to
another can be achieved by sending quantum systems, or by quantum
teleportation.  We have proposed a simple teleportation-based
protocol which allows any quantum operation to be performed upon
$N$ separated, identical quantum systems. Teleportation requires
the transmission of classical information and existence of
entanglement shared between the sending and receiving stations. In
the case of $N=2$, one particular class of operations, namely
those equivalent, up to bipartite local unitary transformations,
to the SWAP operation, permits either the minimum classical
communication or entanglement resources required to perform any
operation to be `recovered' for other tasks.  These operations may
be regarded as the most inseparable operations for $N=2$.

For $N>2$, the situation is more interesting.  For $N{\ge}4$, no
operation can establish the entanglement or be used to communicate
the information necessary to perform any operation.   Whether or
not this is also the case for $N=3$ is currently unknown.  For all
$N$ we have determined the maximum total amount of entanglement
that can be established, and the maximum total number of classical
bits that can be communicated, by any operation. Permutation
operations attain these limits, which are also the minimum
resources required to carry out these specific operations.

We have also examined the problem of finding the minimum resources
required to perform an arbitrary operation.  The scenario we
considered was one where each pair of laboratories shares a
certain amount of entanglement, and can communicate a certain
number of classical bits to each other.  The problem we addressed
was: what are the minimum values of the total entanglement and
communication required to carry out an a priori unknown operation,
that is, unknown prior to the entanglement and communication
resources being set up?

For even $N$, we have found these minimum resources exactly, and
these can be used to perform an arbitrary operation using
teleportation.  We arrived at these bounds using a technique we
refer to as graph symmetrisation. We have shown that the
teleportation protocol is optimal for even {\em N}.  Whether or
not it is also optimal for odd {\em N} is an important outstanding
problem.  We have shown, in the even case, that the optimality of
the teleportation protocol can be reinterpreted as coming from a
restriction on the extent to which expendable resources can be
transferred from one pair of laboratories to another.  In
particular, for any amount of resources transferred, at least as
much are irrevocably lost.  The assumption that this restriction
always holds leads to tighter bounds on the resources required to
carry out an arbitrary operation on an odd number of qubits. These
bounds imply that if, in the one shot scenario, resources can only
be consumed in integer amounts, then the teleportation protocol is
optimal, for all $N>3$ for entanglement and for all ${\em
N}{\geq}12$ for communication also.  One clear conclusion from our
work is that the case of $N=3$ is of particular interest, since
many of our results which apply to all other $N{\neq}2$ have not
been established for this case.  It could be that graph-theoretic
techniques are not suitable for analysing the 3 qubit case, and
that other tools must be employed.
\section*{Acknowledgements.}
\renewcommand{\theequation}{7.\arabic{equation}}
\setcounter{equation}{0}

We would like to thank Sandu Popescu, Noah Linden, Osamu Hirota and Masahide Sasaki for
interesting discussions.  We also wish to thank John Vaccaro for help with some of the
figures. Part of this work was carried out at the Japanese Ministry of Posts and
Telecommunications Communications Research Laboratory, Tokyo, and we would like to thank
Masayuki Izutsu for his hospitality.  This work was funded by the UK Engineering and
Physical Sciences Research Council, and by the British Council.

\end{document}